\documentclass[aps,prl,showpacs,twocolumn,superscriptaddress,showpacs,groupedaddress,nofootinbib]{revtex4-1}  

\usepackage{graphicx}  
\usepackage{dcolumn}   
\usepackage{bm}        
\usepackage{amssymb}   
\usepackage{graphicx,amssymb,amsfonts,amsmath,amssymb,amscd,amstext}
\usepackage[colorlinks=true,citecolor=magenta, linkcolor=blue]{hyperref}

\pdfoutput=1
\def \be {\begin{equation}}
\def \ee {\end{equation}}
\def \bea {\begin{eqnarray}}
\def \eea {\end{eqnarray}}
\def \nn {\nonumber}
\def \la {\langle}
\def \ra {\rangle}

\def \b {\beta}

\def \m {\mu}
\def \n {\nu}

\def\frac#1#2{{#1\over #2}}

\def\tr{\operatorname{tr}}

\newcommand{\oBox}{\mathring{\Box}}
\newcommand{\oD}{\mathring{{\rm \nabla}}}

\newcommand{\oR}{\mathring{R}}

\begin{document}

\begin{flushright}
YITP-SB-16-43
\end{flushright}

\title{
Boundary Fluctuations and A Reduction Entropy}
\author{Christopher Herzog and Kuo-Wei Huang}

\affiliation{\vspace{0.1cm}
Department of Physics and Astronomy, \\
C. N. Yang Institute for Theoretical Physics,\\
Stony Brook University, Stony Brook, NY 11794, USA}

\fontsize{10pt}{11.2pt}\selectfont
\begin{abstract}

The boundary Weyl anomalies live on a codimension-1 boundary, $\partial {\cal M}$. 
The entanglement entropy originates from infinite correlations on both sides of 
a codimension-2 surface, $\Sigma$.  Motivated to have a further understanding of the boundary effects, 
we introduce a notion of reduction entropy, which, guided by thermodynamics,  is a combination of 
the boundary effective action and the boundary stress tensor defined by allowing the metric on $\partial {\cal M}$ 
to fluctuate.  We discuss how a reduction might be performed  so that the reduction entropy 
reproduces the entanglement structure.

\end{abstract}
\maketitle
 
It is believed that a better understanding of black hole entropy 
\cite{BJ, HW}
can be achieved by a deeper understanding of the notion of entropy itself, even in flat spacetime \cite{Wald:1999vt}. 
The concept of entanglement entropy (EE), which serves as a measure of the information lost in correlations on either side of a boundary, 
is widely argued to be a key source of black hole entropy \cite{BKLS, Srednicki:1993im}. Assuming the Hilbert 
space can be factorized into two spatial regions, $A$ and $B$, one defines the EE as 
\bea
S_{\rm{EE}}=- \tr (\rho_A \ln \rho_A) \ ,
\eea
where the reduced density matrix, $\rho_A=\tr_B \rho$, is obtained by tracing over the degrees of 
freedom in the complementary region $B$; $\rho=|\Psi\ra \la \Psi |$ is the full density matrix constructed from a pure state. 

The standard method to compute the EE is the replica method \cite{Holzhey:1994we,  Calabrese:2004eu, Casini:2009sr}, where 
the path integral is performed on an $n$-fold cover of the background geometry with a conical singularity being produced. In this work we  focus on the universal EE for $d=4$ conformal field theories (CFTs) in flat space with an entangling curved surface $\Sigma$. 
The flat space EE obtained using the conical method \cite{Solodukhin:2008dh, Fursaev:2013fta} is given by
\bea
\label{solod}
S_{\rm{EE}}= - {1\over 2 \pi}\int_{\Sigma} \Big(a R_{\Sigma}+c \tr \hat k^2\Big) \ln  ({l\over \delta}) + \textrm{(non-universal)} \ , 
\eea
with
\bea
R_{\Sigma}= \sum^2_{a=1} (k_a^2-\tr k_a^2) \ ,~~ \tr \hat k^2=\sum^2_{a=1}(\tr k_a^2-{1\over 2} k_a^2) \ , 
\eea  
where $a=(1,2)$ represent the coordinates normal to $\Sigma$. 
The non-universal pieces depend on the regularization scheme; $a$ and $c$ are 
central charges. 
Denoting $\gamma_{i  j}$ as the metric on the codimension-2 manifold  $\Sigma$,  the traceless part of the extrinsic curvature 
is $\hat k_{ i  j}\equiv k_{ i j}-{k\over 2} \gamma_{ i j}$, which transforms covariantly under Weyl transformation; $R_{\Sigma}$ 
is the intrinsic Ricci scalar on $\Sigma$.  

Motivated by the holographic computation of the EE  \cite{Ryu:2006bv}, a field theory method was developed 
in \cite{Casini:2011kv}, which shows that employing a conformal transformation allows one to map the EE (restricted for 
a spherical entangling surface, where only the $a$-charge contributes) in CFTs to the ordinary thermodynamical entropy in 
certain curved spaces. This approach however generates a subtle issue related to the boundary effects; these authors found 
a mismatch when comparing the thermal entropy with the universal EE.  The resolution was recently given by \cite{Herzog:2015ioa},  emphasizing 
the importance of boundary terms in the conformal  anomaly.  An  interpretation of \cite{Herzog:2015ioa} is that the universal EE  
can be viewed as a purely boundary effect:
\bea
\label{HHJ}
S_{\rm{EE, ball}}= - \widetilde W[\delta_{\mu\nu}]+ \textrm{(non-universal)} \ ,
\eea
where $\widetilde W[\delta_{\mu\nu}]$ is the $a$-type anomaly effective action with a boundary term, whose 
expression will be given in the next section, evaluated in $flat$ space where only the boundary term contributes. 
A moral of the computation of \cite{Herzog:2015ioa} is that the universal structure of the EE is already dictated 
by flat geometry, and the conformal mapping seems somehow unnecessary.  

The motivation of this work is to generalize \eqref{HHJ}, going beyond the spherical surface restriction. We 
would like to see if the complete universal structure of the EE can be re-captured from boundary anomalies, including 
the $c$-charge contribution, directly from the flat space data. Moreover, we wish not to adopt the conformal mapping or the 
replica method, but simply to rely on the effective action with boundary terms. In other words, we are interested in finding 
a new way to compute the EE. To achieve this goal, there are two immediate challenges. First, the universal 
contribution to the EE comes from a codimension-2 surface while the boundary anomalies live on a codimension-1 manifold; and 
second, the boundary effective action, as we will discuss more later, related to the $c$-charge vanishes in the flat limit. 

We will overcome the first challenge by adopting a metric near $\Sigma$, which allows us to perform a reduction 
sending configurations from $\partial {\cal M}$ to $\Sigma$. We suggest that the second issue can be resolved by 
allowing the metric on $\partial {\cal M}$ to fluctuate: $\delta g_{\mu\nu}|_{\partial \cal{M}}\neq0$. There then exists 
the boundary stress tensor contribution, even in the flat limit. Guided by thermodynamics, we will introduce a notion of 
reduction entropy (RE), which is a combination of the boundary effective action and the boundary stress tensor. Our main result 
is to show how the RE reproduces the EE structure and  therefore conjecture the relation $\text{RE=EE}$ might apply more generally. The discussion of the details we shall leave in the main text. Let us start with a brief review on boundary anomalies.

{\it{Boundary Terms of Conformal Anomaly:}}  
Consider $d=4$ CFTs embedded in a curved spacetime $\cal M$ with a smooth boundary $\partial {\cal M}$. 
The theory can be characterized by the Weyl anomaly.   
The  classification based on the Wess-Zumino consistency \cite{WZ} was presented in \cite{Herzog:2015ioa}. 
(See \cite{a1, Deser:1993yx} for the classification of the bulk anomaly.) 
Denoting the induced metric as $h_{\mu\nu}=g_{\mu\nu}-n_\mu n_\nu$ with $n_\mu$ being a unit, outward normal 
vector to $\partial {\cal M}$, the anomaly is given by  
\bea
\label{4dtrace} 
&&\la T^\mu_\mu\ra
=
{1\over 16 \pi^2} \Big( c W_{\mu\nu\lambda\rho}^2- a E_4\Big)\nn\\
&&+{\delta(x_{\perp})\over 16 \pi^2}  \Big(a E^{\rm{(bry)}}_4-b_1 \tr\hat{K}^3-b_2
 h^{AB}\hat K^{CD} W_{ACBD} \Big) \ ,
\eea  
where $E_4$ is the $d=4$ Euler density 
and $ W_{\mu\nu\lambda\rho}$ is the Weyl tensor.   
 $\delta(x_{\perp})$ is a Dirac delta function with support on the boundary; indices $A,B,C...$ represent boundary coordinates.
The Chern-Simons-like boundary term of the Euler characteristic reads \cite{review} 
\be 
\label{q4} 
E^{\rm{(bry)}}_4
= - 4 \delta^{ABC}_{DEF}~
K^{D}_{A}\left({1\over 2} R^{EF}_{~~~BC} +
{2\over 3} K^{E}_{B} K^{F}_{C} \right) \ , 
\ee  
which is used to supplement $E_4$ to preserve the topological
invariance.   Boundary $b$-type anomalies, with charges $b_1$ and $b_2$, are defined through the traceless 
part of the extrinsic curvature, $\hat{K}_{AB}\equiv K_{AB} - \frac{K}{3}h_{AB}$, which transforms covariantly under  
Weyl scaling.  The $b_1$-and $b_2$-type were pointed out first in
\cite{b1} and \cite{b2}, respectively.  

It is convenient to foliate the spacetime with hypersurfaces labelled by $r$ and adopt the Gaussian normal coordinates. 
The metric is given by $ ds^2= dr^2+ h_{AB}(r,x) dx^A dx^B$.
Using the standard Gauss-Codazzi and Ricci relations, we convert bulk variables into boundary variables and write 
\bea 
&&E^{\rm{(bry)}}_4=  4\left( \frac{2}{3}\text{tr}K^3 -
K \tr K^2+\frac{1}{3}K^3\right)+8 K^{AB}
\mathring{E}_{AB} \ , \\
&&\tr\hat{K}^3=\tr K^3-K \tr K^2+{2\over 9} K^3 \ , \\
&&h^{AB}\hat K^{CD} W_{ACBD} 
= {1\over 6} K^3-{5\over 6} K \tr K^2
+ {1\over 2} K^{AB} \partial_r K_{AB}\nn\\
&&~~~~~~~~~~~~~~~~~~~~~~~~~~~-{1\over 6} K \partial_r K+{1\over 2}  K^{AB} C_{AB} \ , 
\eea  
where we have denoted $\partial_r = n^\mu \partial_\mu$; $\mathring{E}_{AB}$ is the boundary Einstein tensor 
and $C_{AB}=\oR_{AB}- {\oR \over 3} h_{AB}$ is the trace-free part of the 3-dimensional Ricci tensor. ($\oR_{AB}$/$\oR$ denotes the boundary Ricci tensor/scalar.) 

It was recently conjectured that the $b_2$-charge is related to the bulk c-charge by $b_2=8c$.  Using  the heat kernel method, 
ref.\  \cite{Fursaev:2015wpa} confirmed this relation for free fields of spin 0, 1/2, and 1;   
an argument for this relation based on the variational method was given by \cite{Solodukhin:2015eca}.   

A natural question then, which we tentatively answer in the affirmative, is 
if one can recover the $c$-type EE from this $b_2$ boundary anomaly.  
Note that the $b_2$ anomaly \eqref{4dtrace} vanishes in flat space, while the $c$ contribution to the EE \eqref{solod} requires only a curved boundary.
On the other hand, so far there is no indication that the $b_1$ anomaly, which does not vanish in flat space and which depends on boundary conditions, will contribute to the EE. 
We will find that $b_1$ contributes to the entropy in our approach and the known universal EE structure is obtained only when excluding this boundary-condition-dependent charge.
(See \cite{Hawking:1995fd, Barvinsky:1995dp} for earlier discussion of the surface term of the Einstein-Hilbert action and black hole entropy.)


{\it{Response from the Boundary:}} 
Let $W$ be the effective action including boundary terms. The stress tensor in Euclidean space is defined by  $\la T^{\mu\nu}\ra\equiv- {2\over \sqrt{g}} {\delta {{W}} \over \delta g_{\mu\nu}}$. 
We are interested in the anomaly part of the action, denoted as  $\widetilde W$. 
In the dimensional regularization the anomaly effective action for $d=4$ CFTs is essentially given by multiplying the anomaly 
evaluated in $4\to 4+\epsilon$ dimensions by an overall factor ${\mu^{\epsilon}\over \epsilon}$, where $\mu$ stands for a mass scale. 
(To write the effective action in a more  precise manner, one introduces additional vierbeins to construct curvature 
tensors moving away from the physical dimensions; see for instance \cite{Herzog:2013ed, Huang:2016rol} for related discussion.) The log contribution comes
from the expansion ${\mu^{\epsilon}\over \epsilon} \to {1\over \epsilon}+ {\ln \mu}+\cal O(\epsilon)$.  
Hence, we focus on the variation in the physical dimensions.  One should distinguish the divergent part of the stress tensor from the finite part; the effective action is a divergent quantity but the anomaly is a finite
quantity obtained by tracing the finite part of the stress tensor. The divergent part of the variational response contributes to the universal entropy. We denote $\la T^{\mu\nu}\ra= {\mu^{\epsilon}\over \epsilon} \la t^{\mu\nu}\ra$ where  $\la t^{\mu\nu}\ra$ is obtained by varying the integrated anomaly 
with respect to the metric.

It is  useful to adopt the following expression for the EE in $d=4$ CFTs in flat space:
\bea
\label{refined}
\mu {\partial S_{\rm{EE}}\over \partial \mu }
=  c'{{\text{Area} (\Sigma)}\over {(l/\delta)}^2}- {1\over 2 \pi}\int_{\Sigma} \Big(a R_{\Sigma}+c \tr \hat k^2\Big)  \ ,
\eea
where $c'$ is a cut-off-dependent constant; \text{Area}($\Sigma$) is the magnitude of the entangling surface's area. In what follows, we take ${\ln \mu} \to \ln {(l/\delta)}$ where a
dimensional scale $l$ is inserted to have a dimensionless
argument. The quantities we will be computing, the integrated anomaly and $\la t^{\mu\nu}\ra$,  correspond to the contributions to the right-hand-side of \eqref{refined}.  By integrating over \eqref{refined} we have, up to the finite piece, that
\bea
S_{\rm{EE}}
= -{c'\over 2}{{\text{Area} (\Sigma)}\over {(l/\delta)}^2}- {1\over 2 \pi}\int_{\Sigma} \Big(a R_{\Sigma}+c \tr \hat k^2\Big)  \ln ({l\over\delta}) \ .
\eea 
The first term, the area-law of EE, is sometimes dropped in the literature since the coefficient $c'$ depends on the way one determines the cut-off. However, we would like to emphasize that the $a$-charge does not contribute to $c'$ while $c$-charge does contribute to $c'$. (See, for instance, eq(4.28-29) in \cite{Ryu:2006ef} for the discussion.)  In general, there could be non-anomalous contributions to the $1/r^2$ term as well.
Having this in mind, we would like to see  if we can also obtain an additional power-law divergence from the $c$-charge related anomaly action.  

Allowing the boundary metric to fluctuate gives the boundary stress tensor.    The Gaussian normal coordinate is adopted 
after performing the variation. We fix $\delta g_{nA}=0$ where $n$ is the normal-direction. After performing the variation 
we take the flat limit so that the bulk contributions are removed. The shape of the boundary remains generally curved.
Assuming the boundary is smooth and compact, we perform integration by parts along $\partial {\cal M}$ using the 
covariant derivative compatible with the boundary metric. We denote $\oD_A$ as the boundary covariant derivative 
and $\oBox= \oD_A \oD^A$; $D_n= n^\m D_\m$ where $D_\mu$ is the bulk covariant derivative.   The computation 
is straightforward but tedious in details; here we simply state the final results.  

For $a$-type, we obtain  
\bea
\label{at}
&& \lim_{g_{\mu\nu}\to \delta_{\m\n}}{1\over \sqrt{h}}\delta \int_{\partial {\cal M}}E^{\rm{(bry)}}_4\nn\\
&=& 4\int_{\partial M}  \Big( K^{AB}(K^2 -\tr K^2)+2K^{AC}( K_{C}^DK_{D}^{B}-K K_{C}^B) \nn\\
&&~~~~~~~~-  {2\over 3}h^{AB}( \tr K^3- {3\over 2}K \tr K^2+{1\over 2} K^3)\Big) \delta g_{AB}\ .
\eea  
Note the normal-normal component and the contribution $\sim \partial_n \delta g_{AB}$ vanish. 
In curved-space, the contribution $ \sim \partial_n \delta g_{AB}$ gets cancelled by integrating the bulk action by parts. 
Moreover, we observe that the stress tensor obtained from the boundary variation \eqref{at} can be written as $t^A_B\sim \delta^{ACDE}_{BFGH} K^F_C K^G_D K^H_E$, which vanishes identically for any $d=3$ boundary, because of the 4 totally-antisymmetic indices. This in fact simply reflects the topological nature of the Euler characteristic, even in the flat limit.

For the $b_2$-type, we obtain
\bea
\label{hkw}
&&\lim_{g_{\mu\nu}\to \delta_{\m\n}}{1\over \sqrt{h}}\delta   \int_{\partial M}   h^{AB}\hat K^{CD} W_{ACBD} \nn\\
&=&  
 \int_{\partial M} \Big(A^{AB}\delta g_{AB}+
B^{AB}D_n \delta g_{AB}-
C \delta g_{nn}\Big)\nn\\
&&-{1\over 4}\int_{\partial M}\Big(\tr \hat K^2  \partial_n  \delta g_{nn} -\hat K^{AB}  \partial_n (D_n \delta g_{AB})\Big) \ ,
\eea   
with
\bea
A^{AB}&=&   {1\over 6}K^{AB}(K^2 -\tr K^2)+ {1\over 2} K^{AC} (K_{C}^DK_{D}^{B}-K K_{C}^B)\nn\\
&&~~~~+ {1\over 12}(\oBox K^{AB}- h^{AB} \oBox  K)  \ , \\
B^{AB}&=&{1\over 6} h^{AB} (K^2- {3\over 2} \tr K^2) +  {1\over 2} K^A_{C}K^{BC}- {5K \over 12} K^{AB} \ , \nn\\ 
C&=&{1\over 2}\tr K^3
-{2 K\over 3} \tr K^2+{1\over 6} K^3
-{1\over 6}\oBox K\ .
\eea  
This result  implies there are normal derivatives of the metric variation contributions left over on the 
boundary, even in the flat limit. This means the approach of \cite{Solodukhin:2015eca}, which derives $b_2=8c$ 
as a consequence of the variational principle, in fact requires fixing some boundary geometry as a boundary condition (BC).
In general, the anomaly effective action however does not need a well-posed variational principle and the choice of a 
BC would depend on the precise theory one is interested in. (Note we are writing an effective action by  integrating 
out all the dynamical fields; the metric left is an external field.) One of our main tasks therefore is to determine what BC is 
natural for recovering the entanglement structure. We would like to take the viewpoint that the EE boundary is not really a 
physical barrier, and the BC imposed for an EE-related computation could be different from that in other considerations; we 
will discuss more about this point after we introduce the reduction entropy in the next section. 

Finally, for the $b_1$-anomaly we have
\bea
\label{k3t}
{1\over \sqrt{h}}\delta   \int_{\partial M} \tr \hat K^3&=& \int_{\partial \cal{M}} \Big(X^{AB} \delta g_{AB}+Y \delta g_{nn}\Big)\nn\\
&&+ \int_{\partial \cal{M}} Z^{AB} D_n \delta g_{AB}  \ , 
\eea   
with
\bea
X^{AB}&=&{h^{AB}\over 2} \tr\hat{K}^3\ ,~~Y=-{3\over 2}  \tr \hat K^3\ , \\
Z^{AB}&=&{h^{AB}\over 3} (K^2  -{3\over 2}  \tr K^2) +  {3 \over 2}  K^{AC} K^B_{C} -K K^{AB} \ .   
\eea

{\it{Reduction and Entropy:}}  
We aim to relate the structure on $\partial \cal M$ to that of $\Sigma$ through a notion of reduction entropy.
Our basic picture (figure-1) is to first thicken $\Sigma$ by putting a circle with a radius $r$ around each point on $\Sigma$; the 
resulting tube-like manifold is referred to as $\partial \cal M$ on which the boundary anomalies live. Having the information (the 
boundary effective action and the boundary stress tensor) localized on $\partial {\cal M}$, we want to see how these 
configurations contribute when being projected on $\Sigma$. 
\begin{center}
\includegraphics[scale=0.16]{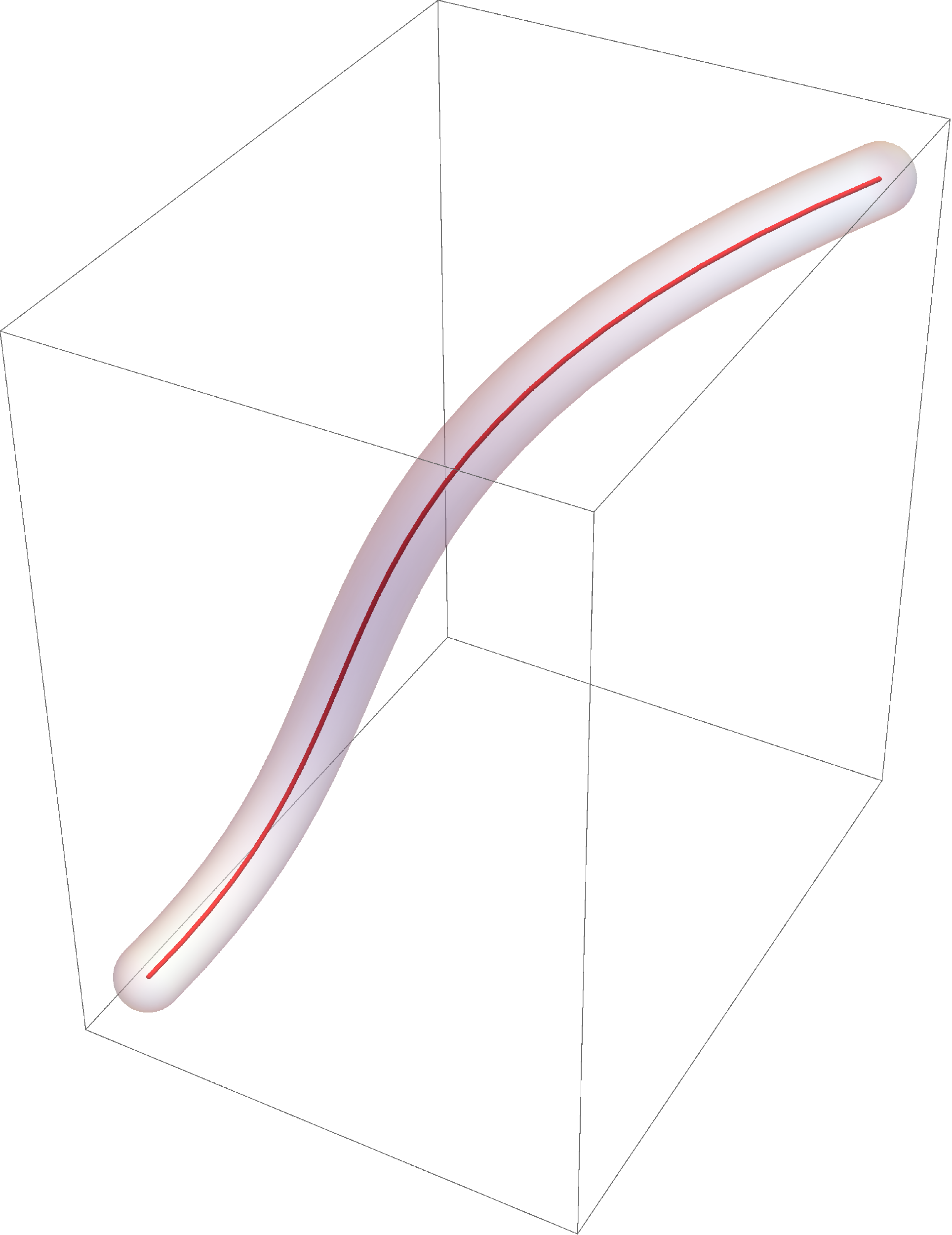}
\\
Figure 1: The codimension-2 surface $\Sigma$ (red line) is thickened to define a codimension-1 boundary 
$\partial {\cal M}$ (tube), where the boundary metric is allowed to fluctuate.  A reduction is performed by 
sending configurations evaluated on $\partial {\cal M}$ back to $\Sigma$.
\end{center}
 
To perform the reduction, we adopt the metric in the vicinity 
of $\Sigma$. We refer the reader to the appendix B in \cite{Rosenhaus:2014woa} (see also \cite{Lewkowycz:2013nqa, Fursaev:2013fta}) for the detailed construction. 
Performing a Wick rotation to Euclidean space, the metric moving away from $\Sigma$ to the second 
order in the distance is given by 
\bea
\label{near}
&&ds^2= \delta_{ab} dx^a dx^b+ A_i \epsilon_{ab} x^a dx^b dy^i+\Big[\gamma_{ij}+2 k^{(a)}_{ij} x^a\nn\\
&&+x^a x^b(\delta_{ab}A_i A_j+k^{(a)}_{i m} k^{(b)m}_{j})\Big]dy^i dy^j+{\cal O}(x^3)\ ,
\eea 
where $\{x^a\}_{a=1,2}$ denote the 2-dimensional transverse spaces and  $\{y^i\}_{i=1,2}$ are coordinates 
parametrizing $\Sigma$, which is located at $x^a=0$; $\epsilon_{ac}$ is the volume form of the transverse space and $\gamma_{ij}$ is 
the corresponding induced metric on the codimension-2 surface.
The gauge field $A_i= {1\over 2}\epsilon^{ac}\partial_a g_{ic}|_{\Sigma}$ is a Kaluza-Klein-like field associated with dimensional 
reduction on $\Sigma$; this gauge field does not contribute upon reduction in our later computation.  (In \eqref{near} we 
have set the background metric to be flat; the metric computed in \cite{Rosenhaus:2014woa} contains curved-space corrections.) 
We next make a transformation to polar coordinates by letting $x^a=r(\cos \theta,\sin \theta)$. The metric becomes
\bea
\label{rsg}
&&ds^2 = dr^2+ r^2 d\theta^2+ 2 r^2 A_i d\theta dy^i+ \Big[\gamma_{ij}+2r\cos\theta k^{(1)}_{ij}\nn\\
&&+2r\sin\theta k^{(2)}_{ij} + r^2 \Big(A_i A_j+\cos^2\theta k^{(1)}_{im} k^{(1) m}_j+ \sin^2\theta k^{(2)}_{im} k^{(2) m}_j
\nn\\
&&~~~~~~~+ \sin(2\theta) k^{(1)}_{im} k^{(2) m}_j\Big)\Big] dy^i dy^j +{\cal O}(x^3)\ . 
\eea
We can use this metric to write the codimension-1 extrinsic
curvature, $K_{AB}$, as a function of the codimension-2 extrinsic curvature, $k^{(a)}_{ij}$:  $K_{AB}\to K_{AB} (k^{(a)}_{ij}, r ,\theta).$ 

We next define a notion of the entropy for this reduction picture. 
Note the polar coordinates we adopt naturally introduce a temperature defined as the inverse of the periodicity $2\pi$. 
But a consequence of adopting the metric \eqref{rsg} is that we are dealing with non-static configurations. 
For instance, $\la T^\theta_\theta\ra$ has explicit $\theta$-dependence with periodicity $2\pi$.  
We consider a natural extension by integrating the $\theta$ (time) variable and define the following notion as the ``reduction entropy" (RE):
\bea
\label{REnew}
S_{\rm{RE}}= \lim_{\partial {\cal M} \to \Sigma} \Big(-\widetilde W+\int_{ {\cal M} } ({{\cal E}}+{{\cal P}})\Big) \ ,
\eea
where $\partial {\cal M} \to \Sigma$ stands for a reduction process.  $\widetilde W$ is the effective action with 
boundary terms. 
The energy density is ${\cal E}\equiv - \la T^\theta_\theta \ra$ and the pressure is interpreted here as the normal-normal component of the stress tensor, ${{\cal P}}= \la T^r_r \ra$. We are largely guided by the  thermodynamical entropy in the  ensemble maintaining constant temperature and pressure; the corresponding entropy reads $S=-W+\beta (\la H\ra + P \la V \ra)$ with $\b=1/T$. 
The reduction is performed by integrating $\theta$ from $0$ to $2\pi$ and taking $r\to 0$ to pick up the contribution localized on $\Sigma$. 
In flat space, all bulk contributions are removed. 
(We find that if instead making an analogy with the canonical ensemble, $S=-W+ \beta \la H\ra$, the c-type EE structure can not be fully recovered and the resulting RE is not Weyl invariant. See the appendix for the discussion. Note that  $\la T^r_r \ra$ is non-zero only for $b_1$-and $b_2$-type actions.)

Let us first consider the simplest case: entropy in $d=2$ CFTs. The anomaly effective action with a boundary is given by
\be
\widetilde W=-{\mu^{\epsilon}\over \epsilon} {c_2\over 24 \pi}\Big( \int_{\cal M} R +2  \int_{\partial{\cal M}} K\Big), ~~\epsilon=d-2 \ ,
\ee 
where $c_2$ stands for the central charge in $d=2$. We focus on the $d\to2$ divergent contribution. 
The metric variation gives the boundary stress tensor $t^A_B  \sim (K^A_B- h^A_B K)= \delta^{AC}_{BD} K^D_C$, which vanishes identically for any $d=1$ boundary.
Thus, the pressure and the energy do not contribute to $d=2$ RE. 
The partition function in flat space is determined by the boundary term. Note here $\Sigma$ represents two end-points of 
an entangling interval. We obtain (up to non-universal pieces)
\bea
S_{\rm{RE}}&=&-\lim_{\partial {\cal M} \to \Sigma} \widetilde W
= {c_2\over 3 } \ln  {({l\over \delta})} =  S_{\rm{EE}} \ ,
\eea which is the classic universal EE in $d=2$ CFTs \cite{Holzhey:1994we}. 

We next turn to the $d=4$ $a$-type contribution. There is no $a$-type boundary stress tensor contribution.
Thus,
\bea
&&\lim_{\partial {\cal M} \to \Sigma}  \int_{{\cal M}} {\la t^\theta_\theta\ra}^{(a)}=\lim_{\partial {\cal M} \to \Sigma}  \int_{ {\cal M}} {\la t^r_r\ra}^{(a)}=0\ .
\eea
The $a$-type RE then solely comes from the partition function, with only the boundary term surviving in flat space.  
Performing the reduction, we find, up to non-universal terms, that
\bea
S^{(a)}_{\rm{RE}}&=&-\lim_{\partial {\cal M} \to \Sigma} \widetilde W^{(a)}=-\Big({a\over 2 \pi } \int_{ \Sigma} R_{\Sigma} \Big)  \ln  {({l\over \delta})}\ . 
\eea 
This recovers the $a$-type EE \eqref{solod}. This expression generalizes the earlier result \eqref{HHJ}, which is restricted to a spherical surface. 
The boundary stress tensor was ignored in \cite{Herzog:2015ioa}. However, as shown in the present approach, even if one 
restores the boundary stress tensor, the result of \cite{Herzog:2015ioa} remains valid. 

The more challenging part is to recover also the $c$-type EE. Having this $b_2=8c$ relation, we next consider the $b_2$-charge 
contribution to RE. In this case, the RE is given instead only by the boundary stress tensor, since the action simply vanishes in the flat  limit.  
We should now discuss our choice of BC. The standard way to determine the corresponding boundary stress tensor from 
the result \eqref{hkw} is by imposing certain BCs removing all normal derivatives of the metric variation.  However, imposing any 
Neumann-like BC might not be natural for this entanglement computation because the EE surface (and $\partial {\cal M}$ in the  RE picture) should not be viewed as a real boundary.  
Thus, in this particular computation, we chose $not$ to impose any BC or any constraint on the boundary geometry.  
The resulting $b_2$-boundary stress tensor then contains some normal derivatives of the Dirac delta function left over on the boundary.

Let us discuss how these delta functions contribute to RE. We first adopt the following expression in the Gaussian normal coordinates:
\bea
D_r \delta g_{AB}= \partial_r \delta g_{AB}- K_A^C \delta g_{BC}- K_B^C \delta g_{AC}\ .
\eea The last two terms contribute to the stress tensor in the standard manner. 
We also have
\bea
\label{ddg}
&& \lim_{g_{\mu\nu}\to \delta_{\mu\nu}}\partial_r (D_r \delta g_{AB})=\partial_r \partial_r \delta g_{AB}+ K_A^D K^C_D \delta g_{BC}\nn\\
&&~~~~+K_B^D K^C_D \delta g_{AC}- K_A^C \partial_r\delta g_{BC}-K_B^C \partial_r\delta g_{AC} \ ,
\eea 
where we have used that in flat space $\partial_r K^B _A= - K_A^C K^B_{C}$. 
Including all the contributions, the $b_2$-type stress tensor reads
\bea
\label{full}
&&\la t^{AB} \ra^{(b_2)}= -{b_2\over 8 \pi^2}\Big({A^{AB}} + \Delta_1^{AB} + \Delta_2^{AB}\Big)\delta (r-r')
\eea
with
\bea
 \Delta_1^{AB} &=&{B^{AB}}\partial_r -2 B^{AC} K^B_{C} \ , \\
 \Delta_2^{AB} &=& {\hat K^{AB} \over 4} \partial^2_r  
-{\hat K^{AC} K_C^B \over 2} \partial_r   +{ \hat K^{A}_C\over 2} K^B_D  K^{CD}\ ,
\eea  
where $r'$ denotes the location of $\partial {\cal M}$; by sending $r'\to 0$, we are performing the reduction back to the location of $\Sigma$.  We also have
\bea
\label{pb2}
\la t^{rr} \ra^{(b_2)}= {b_2\over 8 \pi^2} \Big(C+{1\over 4}\tr \hat K^2 \partial_r\Big) \delta (r-r') \ .
\eea
Notice in the formula \eqref{REnew} we integrate the density over all space. (The bulk stress tensor vanishes in the flat limit.)   
Focusing on the normal-coordinate dependence and letting $f(r)$ be the structure that multiplies the derivative of the delta function, in computing RE we use the property that 
$\int dr  f(r) \partial_r \delta (r-r')  
= - \partial_{r'}  \int dr f(r) \delta (r-r')
= - \partial_{r'} f(r')$ to proceed. 
 Similarly we have $\int dr  f(r) \partial^2_r \delta (r-r')  =  \partial^2_{r'} f(r')$.  Note $f(r)$ includes the measure factor.  We do not perform integration by parts using the normal-derivative on the boundary. 

Performing the reduction we obtain, in flat space,
\bea
\label{mag}
\mu {\partial \over \partial \mu}S^{(b_2=8c)}_{\rm{RE}}&=& \lim_{\partial {\cal M} \to \Sigma}  \int_{{\cal M}} \Big(-{\la t^\theta_\theta\ra}+{\la t^r_r\ra}\Big)^{(b_2=8c)}\nn\\
&=& {c\over 3\pi} {\text{Area}(\Sigma)\over r_{\text{cut-off}}^2} -{c\over 2 \pi } \int_{ \Sigma} \tr \hat k^2 \ ,
\eea 
where we have adopted the conjectured $b_2=8c$ relation.
The first term gives the area-law of the entropy as mentioned in \eqref{refined}.  
(Such a power-law divergence gets intrinsically cancelled in the $a$-type RE.)
The second term of \eqref{mag} reproduces the universal $c$-type EE \eqref{solod}. 
Showing RE=EE from first principles would then provide a proof of $b_2=8c$.

Finally, we shall also discuss the $b_1$ boundary anomaly. Note the action does not vanish in flat space. 
The boundary stress tensor can be read from \eqref{k3t}. We have
\bea
\la t^{AB} \ra^{(b_1)}&=& -{b_1\over 8 \pi^2} \Big(X^{AB}+ {Z^{AB}\partial_r} -2 Z^{AC} K^B_{C}\Big) \delta (r-r') \ , \nn\\
\la t^{rr} \ra^{(b_1)}&=& - {b_1\over 8 \pi^2} Y  \delta (r-r') \ .
\eea  To be consistent, we again do not impose any BC. Adopting the same method discussed before to perform the reduction, we find that there is no power-law divergence and we obtain
\bea
\label{b1RE}
S^{(b_1)}_{\rm{RE}}&=& {b_1\over 48 \pi} \Big(\int_{ \Sigma}R_{\Sigma}\Big) \ln  {({l\over \delta})} \ ,
\eea up to non-universal terms.
We see it contributes like the topological $a$-type EE.  The result \eqref{b1RE} deserves a further understanding.   
It would be of great interest to see how the $b_1$-charge might touch EE.  But an interpretation might be that since $b_1$-charge is sensitive to BCs 
imposed on matter fields \cite{Fursaev:2015wpa}, one might in this sense not regard it as a universal contribution to the entropy. Another possibility, which we do not explore further here, is that $b_1$ might be related to the $c$-charge under certain BCs imposed on matter fields that are suitable in the EE computation; see the appendix for a related viewpoint.

{\it{Conclusion:}}
In this paper we have tried to establish two main messages: First, the boundary geometry related to the anomaly 
is rather rich and of great potential importance. The boundary terms are however largely ignored in the literature 
when constructing a theory containing an entangling surface or a conical singularity; and secondly, it is possible 
to compute EE directly from flat space, without introducing the $n$-fold manifold or performing a 
conformal mapping to a curved space. 
Our computation suggests the following identification:
\bea
\label{REisEE}
S_{\rm{RE}}=S_{\rm{EE}} \ ,
\eea  
up to non-universal pieces. 
(The conjecture \eqref{REisEE} is formulated by omitting the additional boundary-condition-dependent $b_1$ contribution to the RE. The expression \eqref{REisEE} is then independent of boundary conditions imposed on the matter fields.)

There are however considerable questions and puzzles that are worthy of future study. 
The most important one perhaps is to have a  direct understanding why the universal pieces of the RE and EE should match. 
A heuristic argument is that the universal structure due to thermal excitations near a codimension-2 surface is  indistinguishable 
from that due to entanglement.  Let us here list three further topics for investigation:
(1) It would be interesting to restore boundary terms also in the replica approach and see how the boundary terms might interact with the conical singularity.  
(2) It might be possible to search for new BCs for the boundary metric, different from the one considered here, for RE and compare it with EE. 
(3) One should better understand why ${\la t^r_r \ra}$ is important when recovering the EE structure.
\\

{\it{Acknowledgments:}}  
We  thank  H. Casini, D. Fursaev, R. Myers,  D. Page, 
S. Solodukhin and T. Takayanagi for interesting conversations, remarks on boundary anomalies and useful comments. 
This work was supported in part by the National Science Foundation under Grant No. PHY13-16617 and PHY-1620628.
\\

{\it{Appendix:}}  
If we adopt the canonical expression,
\bea
\label{RE}
S_{\rm{RE}}= \lim_{\partial {\cal M} \to \Sigma} \Big(-\widetilde W+\int_{ {\cal M} } {\cal E}\Big) \ ,
\eea 
we instead find 
\bea
\label{can}
\mu {\partial \over \partial \mu} S_{\rm{RE}}&=& - {b_1\over 1 2 \pi} {\text{Area}(\Sigma)\over  r_{\text{cut-off}}^2} +\Big({2b_1+{b_2}\over 16 \pi} \int_{ \Sigma} \tr \hat k^2 \nn\\
&&-{b_2\over 48 \pi} \int_{ \Sigma} \tr k^2 - {a\over 2 \pi } \int_{ \Sigma} R_{\Sigma}\Big)    \ .
\eea The a-type result is untouched. If matching the above expression with the universal EE one has
\bea
2b_1+b_2= - 8c \ , 
\eea 
as a consistency condition. (If $b_2$ is still $8c$, then $b_1=-b_2= -8c$.) It would be nice to see under what kind of BC does this scenario apply. 
However, there is an additional $\sim \int_{ \Sigma} \tr k^2 $ contribution in \eqref{can}.  We 
notice such a non-conformal invariant structure appears in \cite{Astaneh:2014sma} as a potential contribution to EE from the scheme-dependent $\Box R$ anomaly. 
The $\Box R$ term is produced by an $R^2$ effective action; we emphasize that the corresponding action 
is $finite$. It reads $\widetilde W \sim {\mu^\epsilon\over \epsilon} (\epsilon R^2)$. 
(If one instead takes $\widetilde W \sim {\mu^\epsilon\over \epsilon} R^2$, the 
Weyl transformation generates an $R^2$ which violates the Wess-Zumino consistency condition.) 
Therefore, the universal part of the RE is free from this $\Box R$ ambiguity. 
In this sense  RE is more robust than EE, and one might instead ask what scheme used in EE can match with RE. 
It would be interesting to further investigate the potential connection between the boundary anomaly and the $\Box R$  bulk anomaly.

\end{document}